\documentclass[10pt,twocolumn,pre,aps,showpacs,superscriptaddress,showpacs,byrevtex,bibnotes]{revtex4}
\usepackage{amsfonts}
\usepackage{graphics}
\usepackage{amsthm}

\newtheorem{theorem}{Theorem}

\begin{document}

\title{Generalized canonical ensembles and ensemble equivalence}

\author{M. Costeniuc}
\affiliation{Department of Mathematics and Statistics, University of Massachusetts,
Amherst, MA 01003, USA }

\author{R. S. Ellis}
\email{rsellis@math.umass.edu}
\affiliation{Department of Mathematics and Statistics, University of Massachusetts,
Amherst, MA 01003, USA }

\author{H. Touchette}
\email{htouchet@alum.mit.edu}
\affiliation{School of Mathematical Sciences, Queen Mary, University of London,
London E1 4NS, UK}

\author{B. Turkington}
\email{turk@math.umass.edu}
\affiliation{Department of Mathematics and Statistics, University of Massachusetts,
Amherst, MA 01003, USA }

\date{\today}

\begin{abstract}

This paper is a companion article to our previous paper 
(J.~Stat.~Phys.~\textbf{119}, 1283 (2005), cond-mat/0408681),
which introduced a generalized canonical ensemble obtained by multiplying
the usual Boltzmann weight factor $e^{-\beta H}$ of the canonical ensemble
with an exponential factor involving a continuous function $g$ of the
Hamiltonian $H$. We provide here a simplified introduction to our previous
work, focusing now on a number of physical rather than mathematical aspects
of the generalized canonical ensemble. The main result discussed is that,
for suitable choices of $g$, the generalized canonical ensemble reproduces,
in the thermodynamic limit, all the microcanonical equilibrium properties of
the many-body system represented by $H$ even if this system has a nonconcave
microcanonical entropy function. This is something that in general the
standard ($g=0$) canonical ensemble cannot achieve. Thus a virtue of the
generalized canonical ensemble is that it can be made equivalent to the
microcanonical ensemble in cases where the canonical ensemble cannot. The
case of quadratic $g$-functions is discussed in detail; it leads to the
so-called Gaussian ensemble.

\end{abstract}


\maketitle

\section{Introduction}

The study of many-body systems having nonconcave entropy functions has been
an active topic of research for some years now, with fields of study ranging
from nuclear fragmentation processes \cite{agostino2000,gross1997,gross2001},
and phase transitions in general \cite{chomaz2001,gulminelli2002,touchette2005},
to statistical theories of stars formation
\cite{lynden1968,lynden1999,chavanis2002,chavanis22002,chavanis2003,chavanis22003},
as well as statistical theories of fluid turbulence \cite{ellis2000,ellis2002}.
The many different systems covered by these studies share an interesting
particularity: they all have equilibrium properties or \textit{states} 
that are seen in the microcanonical ensemble but not in the canonical ensemble.
Such  \textit{microcanonical nonequivalent states}, as they are called,
directly arise as a result of the nonconcavity of the entropy function,
and can present themselves in many different ways both at the
thermodynamic level (e.g., as negative
values of the heat capacity \cite{lynden1999,thirring1970}) and the level of
general macrostates (e.g., as canonically-unallowed values of the
magnetization \cite{ellis2000,ellis2004}).

The fact that the canonical ensemble misses a part of the microcanonical
ensemble when the entropy function of that latter ensemble is nonconcave can
be understood superficially by noting two mathematical facts:

(i) The free energy function, the basic thermodynamic function of the
canonical ensemble, is an always concave function of the inverse temperature.

(ii) The Legendre(-Fenchel) transform, the mathematical transform that
normally connects the free energy to the entropy, and vice versa, only
yields concave functions.

Taken together, these facts tell us that microcanonical entropy functions
that are nonconcave cannot be expressed as the Legendre(-Fenchel) transform
of the canonical free energy function, for otherwise these entropy functions
would be concave. One should accordingly expect in this case to observe
microcanonical equilibrium properties that have absolutely no equivalent in
the canonical ensemble, since the energy and the temperature should then
cease to be related in a one-to-one fashion, as is the case when the entropy
function is strictly concave. This is indeed what is predicted theoretically
\cite{eyink1993,ellis2000} and what is observed in many systems, including
self-gravitating systems
\cite{lynden1968,lynden1999,chavanis2002,chavanis22002,chavanis2003,chavanis22003},
models of fluid turbulence \cite{ellis2000,ellis2002}, atom clusters
\cite{schmidt2001,gobet2002}, as well as long-range interacting spin models
\cite{dauxois2000,ispolatov2000,barre2001,draw2002,antoni2002,touchette2003,costeniuc2005}
and models of plasmas \cite{smith1990}.

What we present in this paper comes as an attempt to specifically assess the
nonequivalent properties of a system which are seen at equilibrium in the
microcanonical ensemble but not in the canonical ensemble. Obviously, one
way to predict or calculate such properties is to proceed directly from the
microcanonical ensemble. However, given the notorious intractability of
microcanonical calculations
\footnote{The microcanonical ensemble is generally more tedious to
work with than the canonical ensemble, both analytically and
numerically, as the microcanonical ensemble is defined with an
equality constraint on the energy, while the canonical ensemble
involves no such constraint.}, it seems sensible to consider the possibility
of modifying or generalizing the canonical ensemble in the hope that it can
be made equivalent with the microcanonical ensemble while preserving its
analytical and computational tractability. Our aim here is to show how this
idea can be put to work in two steps: first, by presenting the construction
of a generalized canonical ensemble, and, second, by offering proofs of its
equivalence with the microcanonical ensemble. Our generalized canonical
ensemble, it turns out, not only contain the canonical ensemble as a special
case, but also incorporates the so-called Gaussian ensemble proposed some
years ago by Hetherington \cite{heth1987}. The proofs of equivalence that we
present here for the generalized canonical ensemble also apply therefore
to the Gaussian ensemble.

Much of the content of the present paper has been exposed in a previous
paper of ours \cite{costeniuc2004}. The reader will find in that paper a
complete and rigorous mathematical discussion of the generalized canonical
ensemble. The goal of the present paper is to complement this discussion by
presenting it in a less technical way than previously done and by
highlighting a number physical implications of the generalized canonical
ensemble which were not discussed before.

The content of the paper is as follows. In the next section, we review the
theory of nonequivalent ensembles so as to set the notations and the basic
results that we seek to generalize in this paper. This section is also
meant to be a review of the definitions of the microcanonical and canonical
ensembles. In Section~\ref{Sgen}, we then present our generalization of the
canonical ensemble and
give proofs of its equivalence with the microcanonical ensemble for both the
thermodynamic level and the macrostate level of statistical mechanics.
Section~\ref{Sgauss} specializes these considerations to the special case of the
Gaussian ensemble. We briefly comment, finally, on our ongoing work on
applications of the generalized canonical ensemble.

\section{Review of nonequivalent ensembles}

\label{Srev}

We consider, as is usual in statistical mechanics, an $n$-body system with
microstate $\omega \in \Omega _n$ and Hamiltonian $H(\omega )$; $\Omega _n$
is the microstate space. Denoting the mean energy of the system by $h(\omega
)=H(\omega )/n$, we define the microcanonical entropy function of the system
by the usual limit
\begin{equation}
s(u)=\lim_{n\rightarrow \infty }\frac 1n\ln \rho _n(u),
\end{equation}
where
\begin{equation}
\rho _n(u)=\int_{\{\omega \in \Omega _n:h(\omega )=u\}}d\omega =\int_{\Omega
_n}\delta (h(\omega )-u)d\omega
\end{equation}
represents the density of microstates $\omega $ of the system having a mean
energy $h(\omega )$ equal to $u$. As is well-known, $s(u)$ is the basic
function for the microcanonical ensemble from which one calculates the
thermodynamic properties of the system represented by $h(\omega )$ as a
function of its energy $nu$. The analogous function for the canonical
ensemble which is used to predict the thermodynamic behavior of the system
as a function of its temperature $T=(k_B\beta )^{-1}$ is the free energy
function $\varphi (\beta )$. The latter function is taken here to be defined
by the limit
\begin{equation}
\varphi (\beta )=\lim_{n\rightarrow \infty }-\frac 1n\ln Z_n(\beta ),
\label{free1}
\end{equation}
where
\begin{equation}
Z_n(\beta )=\int_{\Omega _n}e^{-n\beta h(\omega )}d\omega
\end{equation}
denotes, as usual, the partition function of the system at inverse
temperature $\beta =(k_BT)^{-1}$.

\begin{figure}[t]
\includegraphics{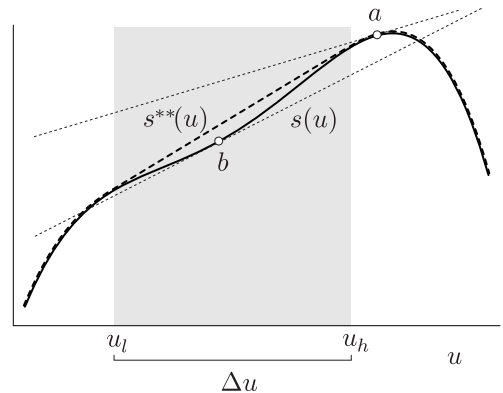}
\caption{Geometric interpretation of supporting lines in relation to the
graph of the microcanonical entropy function $s(u)$ (full line) and its
concave envelope or concave hull $s^{**}(u)$ (dashed line).
The point $a$ in the figure has the property that $s(u)$ admits a
supporting line at $a$; i.e., there exists a line passing through
$(a,s(a))$ that lies above the graph of $s(u)$. In this case, $s(a) =
s^{**}(a)$. The point $b$ in the figure has the property that $s(u)$
admits no supporting line at $b$.  In this case $s(b) \not = s^{**}(b)$.}
\label{figline}
\end{figure}

The entropy and free energy functions are obviously two different functions
that refer to two different physical situations---the first to a closed
system having a fixed energy, the second to an open system in contact with a
heat bath having a fixed inverse temperature. However, these two functions
are not independent. In fact, we only have to rewrite the integral defining
the partition function $Z_n(\beta )$ as an integral over the mean energy
values
\begin{equation}
Z_n(\beta )=\int \rho _n(u)e^{-n\beta u}du
\end{equation}
rather than an integral over $\Omega _n$, and then approximate the resulting
integral using Laplace's method, to see that
\begin{equation}
Z_n(\beta )\approx \exp \left( -n\inf_u\{\beta u-s(u)\}\right)
\end{equation}
with subexponential correction factors in $n$. This application of Laplace's
approximation is quite standard in statistical mechanics and leads us
hitherto to the following important equation:
\begin{equation}
\varphi (\beta )=\inf_u\{\beta u-s(u)\},  \label{lf1}
\end{equation}
which expresses $\varphi (\beta )$ as the \textit{Legendre-Fenchel} (LF)
transform of $s(u)$ \cite{ellis1985,ellis2000}. In convex analysis, the LF
transform is often abbreviated by the notation $\varphi =s^{*}$, and $s^{*}$
in this context is called the \textit{dual} of $s$
\cite{rockafellar1970,ellis1985,ellis2000}. It can be shown that the basic
relationship $\varphi =s^{*}$ holds no matter what shape $s(u)$ has, be it
concave or not~\cite{ellis2000}. Consequently, $\varphi (\beta )$ can always
be calculated from the microcanonical ensemble by first calculating $s(u)$
and then take the LF\ transform of this latter function. That this procedure
always yield the correct free energy function $\varphi (\beta )$ follows
basically from the fact that $\varphi (\beta )$ is an always concave
function of $\beta $ \cite{ellis1985}.

It is the converse process, that is, the attempt of calculating $s(u)$ from
the point of view of the canonical ensemble by calculating the LF transform
of $\varphi (\beta )$ which is problematic. Contrary to $\varphi (\beta )$,
$s(u)$ need not be an always concave function of $u$. This has for
consequence that the double LF\ transform $\varphi ^{*}=(s^{*})^{*}$, which
takes the explicit form
\begin{equation}
\varphi ^{*}(u)=s^{**}(u)=\inf_\beta \{\beta u-\varphi (\beta )\},
\label{alf1}
\end{equation}
may not necessarily yield $s(u)$ since the LF transform of a concave
function, here $\varphi (\beta )$, yields a concave function. At this point,
the key question that we have to ask then is: when does $s^{**}(u)$ equal
$s(u)$?

The answer to this question is provided by the theory of convex functions
\cite{rockafellar1970,ellis2000}, and invokes a concept central to this
theory known as a \textit{supporting line}. This is the subject of the next
theorem which we state without a proof; see Ref.~\cite{ellis2000} for details.

\begin{theorem}
\label{Tsupp}
We say that $s$ admits a supporting line at $u$ if there exists
$\beta $ such that $s(v)\leq s(u)+\beta (v-u)$ for all $v$
\emph{(see Fig.~\ref{figline})}.

\emph{(a)} If $s$ admits a supporting line at $u$, then
\begin{equation}
s(u)=\inf_\beta \{\beta u-\varphi (\beta )\}=s^{**}(u).  \label{lf2}
\end{equation}

\emph{(b)} If $s$ admits no supporting line at $u$, then
\begin{equation}
s(u)\neq \inf_\beta \{\beta u-\varphi (\beta )\}=s^{**}(u).
\end{equation}
\end{theorem}

In the former case where $s$ admits a supporting line, we say that the
microcanonical and canonical ensembles are \textit{thermodynamically
equivalent at} $u$, since then the microcanonical entropy function can be
calculated from the point of view of the canonical ensemble by taking the LF
transform of free energy function. In the opposite case, namely when $s$
does not admit a supporting line, we say that the microcanonical and
canonical ensembles are \textit{thermodynamically nonequivalent at} $u$
\cite{ellis2000,touchette2003,touchette2004}. Note that $s^{**}(u)$ represents in
general the concave envelope or \textit{concave hull} of $s(u)$ which is the
smallest concave functions satisfying $s^{**}(u)\geq s(u)$ for all values of
$u$ in the range of $h$ (see Fig.~\ref{figline}). Hence, $s(u)<s^{**}(u)$
if $s(u)\neq s^{**}(u)$. Note also
that if $s$ is differentiable at $u$, then the slope $\beta $ of its
supporting line, if it has one, has the value $\beta =s^{\prime }(u)$
\cite{ellis2000}.

The nonequivalence of the microcanonical and canonical ensembles can also be
stated alternatively from the point of view of the canonical ensemble as a
definition involving the free energy. All that is required is to use the
fact that the LF transform of a strictly concave, differentiable function
(negative second derivative everywhere) yields a function which is also
strictly concave and differentiable \cite{rockafellar1970}. This is stated
next without proof (see Refs.~\cite{ellis2000,touchette2003,touchette2005}).

\begin{theorem}
\label{Tphi}
Let $\varphi (\beta )$ denote the free energy function defined
in \emph{(\ref{free1})}.

\emph{(a)} If $\varphi $ is differentiable at $\beta $, then
\begin{equation}
s(u_\beta )=\varphi ^{*}(u_\beta )=\beta u_\beta -\varphi (\beta ),
\end{equation}
where $u_\beta =\varphi ^{\prime }(\beta )$ represents the equilibrium value
of $h$ in the canonical ensemble with inverse temperature $\beta $.

\emph{(b)} If $\varphi $ is everywhere differentiable, then $s=\varphi ^{*}$
for all $u$ in the range of $h$.
\end{theorem}

\begin{figure}[t]
\includegraphics{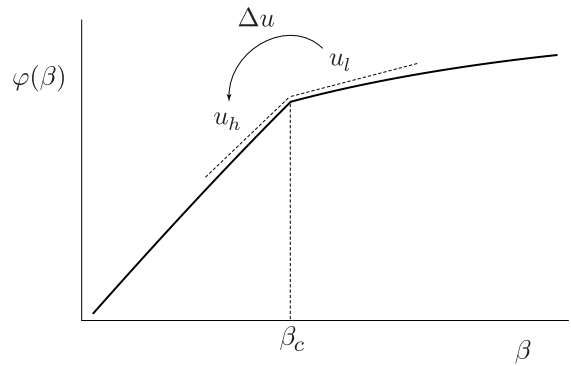}
\caption{Free energy function $\protect\varphi (\protect\beta)$ associated with
the nonconcave entropy function $s(u)$ shown in Fig.~\ref{figline}.
The region of nonconcavity of $s(u)$ is signaled at the level of
$\protect\varphi (\protect\beta)$
by the appearance of a point
$\protect{\beta_c}$ where $\varphi({\protect\beta})$ is nondifferentiable.
$\protect{\beta_c}$ equals
the slope of the affine part of $s^{**}(u)$, while the left-
and right-derivatives of $\varphi$ at $\protect{\beta_c}$
equal $u_h$ and $u_l$, respectively.}
\label{figphi}
\end{figure}

This last result is useful because it pinpoints the precise physical cause
of nonequivalent ensembles, namely, the emergence of first-order phase
transitions in the canonical ensemble, as signaled by nondifferentiable
points of $\varphi (\beta )$. Put simply, but not quite rigorously, there
must be nonequivalence of ensembles whenever the canonical ensemble
undergoes a first-order transition because, in crossing the critical inverse
temperature $\beta _c$ at which $\varphi (\beta )$ is nondifferentiable,
this ensemble skips over an interval of mean energy values that are
accessible within the microcanonical ensemble
\cite{thirring1970,lynden1999,gross1997,gross2001,chomaz2001,gulminelli2002,touchette2006}.
The ``skipped'' interval in this case is precisely given by $(u_l,u_h)$,
where $u_l$ and $u_h$ are the lower and upper values at which we have
thermodynamic nonequivalence of ensembles, that is, at which $s(u)\neq
s^{**}(u)$ (Fig.~\ref{figline}). Going to the canonical ensemble,
it can be shown without too
much difficulties that these boundary values are also such that $u_l=\varphi
^{\prime }(\beta _c+0)$ and $u_h=\varphi ^{\prime }(\beta _c-0)$, where
$\varphi ^{\prime }(\beta _c+0)$ and $\varphi ^{\prime }(\beta _c-0)$ denote
the right- and left-side derivatives of $\varphi $ at $\beta _c$,
respectively (Fig.~\ref{figphi}).
Therefore, from the canonical point of view, the length
$\Delta u=u_h-u_l$ of the nonconcavity interval of $s(u)$ corresponds to the
latent heat of a first-order phase transition.

\section{Generalized canonical ensemble}

\label{Sgen}

We now introduce a new canonical ensemble that, as we will prove, can be
made equivalent with the microcanonical ensemble in cases when the standard
canonical ensemble is not. The construction of this generalized canonical
ensemble follows simply by replacing the Lebesgue measure $d\omega $
entering in the integral of the partition function $Z_n(\beta )$ with the
new measure $e^{-ng(h(\omega ))}d\omega $, where $g(h)$ is a continuous but
otherwise arbitrary function of the mean Hamiltonian $h(\omega )$. Thus,
\begin{equation}
Z_{g,n}(\alpha )=\int_{\Omega _n}e^{-n\alpha h(\omega )-ng(h(\omega
))}d\omega
\end{equation}
represents the partition of our system in the generalized canonical ensemble
with parameter $\alpha $. The corresponding generalized free energy is
\begin{equation}
\varphi _g(\alpha )=\lim_{n\rightarrow \infty }-\frac 1n\ln Z_{g,n}(\alpha ).
\label{genfe1}
\end{equation}
We use at this point the variable $\alpha $ in lieu of $\beta $ in order not
to confuse $\alpha $ with the inverse temperature of the canonical ensemble.

At the level of probabilities, the change of measure $d\omega \rightarrow
e^{-ng(h(\omega ))}d\omega $ leads us naturally to consider the following
probability density:
\begin{equation}
p_{g,\alpha }(\omega )=\frac{e^{-n\alpha h(\omega )-ng(h(\omega ))}}{
Z_{g,n}(\alpha )}
\end{equation}
as defining our generalized canonical ensemble. The choice $g=0$ yields back
obviously the standard canonical ensemble; that is,
\begin{equation}
p_{g=0,\alpha }(\omega )=\frac{e^{-n\alpha h(\omega )}}{Z_n(\alpha )}
\end{equation}
and $\varphi _{g=0}(\alpha )=\varphi (\beta =\alpha )$.

Let us now show how the generalized canonical ensemble can be used to
calculate the microcanonical entropy function. Repeating the steps which led
us to express $\varphi (\beta )$ as the LF transform of $s(u)$, it is
straightforward to derive the following modified LF transform:
\begin{equation}
\varphi _g(\alpha )=\inf_u\{\alpha u+g(u)-s(u)\}
\end{equation}
which, by defining $s_g(u)=s(u)-g(u)$, can be written in the form
\begin{equation}
\varphi _g(\alpha )=\inf_u\{\alpha u-s_g(u)\}.  \label{genlf1}
\end{equation}
This shows that the generalized free energy $\varphi _g(\alpha )$ is the LF
transform of a deformed entropy function $s_g(u)$. This function can be
thought of as representing the entropy function of a generalized
microcanonical ensemble defined by the following modified density of states:
\begin{equation}
\rho _{g,n}(u)=\int_{\Omega _n}\delta (h(\omega )-u)e^{-ng(h(\omega
))}d\omega .
\end{equation}
Note indeed that $\rho _{g,n}(u)=e^{-ng(u)}\rho _n(u)$, so that
\begin{eqnarray}
s_g(u) &=&\lim_{n\rightarrow \infty }\frac 1n\ln \rho _{g,n}(u)  \nonumber \\
&=&-g(u)+\lim_{n\rightarrow \infty }\frac 1n\ln \rho _n(u)  \nonumber \\
&=&s(u)-g(u).  \label{defent1}
\end{eqnarray}

As was the case for standard canonical free energy $\varphi (\beta )$, the
LF\ transform that now relates $\varphi _g(\alpha )$ to the LF transform of
$s_g(u)$ can be shown to be valid for any function $s(u)$ and any choice of
$g$ since $\varphi _g(\alpha )$ is an always concave function of $\alpha $.
However, as before, the reversal of this transform is subjected to a
supporting line condition which now takes effect at the level of $s_g(u)$.
More precisely, if $s_g$ admits a supporting line at $u$, in the sense that
there exists $\alpha $ such that
\begin{equation}
s_g(v)\leq s_g(u)+\alpha (v-u)
\end{equation}
for all $v$, then the transform $\varphi _g^{*}$ yields the correct entropy
function $s_g$ at $u$, that is,
\begin{equation}
s_g(u)=\inf_\alpha \{\alpha u-\varphi _g(\alpha )\}=s_g^{**}(u);
\end{equation}
otherwise $s_g(u)\neq s_g^{**}(u)$. At this point, we only have to use the
fact that $s(u)=s_g(u)+g(u)$ to obtain the following result.

\begin{theorem}
\label{Tgenth}
Let $g(u)$ be a continuous function of $u$ in terms of which
we define $s_g(u)=s(u)-g(u)$.

\emph{(a)} If $s_g$ admits a supporting line at $u$, then
\begin{equation}
s(u)=\inf_\alpha \{\alpha u-\varphi _g(\alpha )\}+g(u).
\end{equation}

\emph{(b)} If $s_g$ does not admit a supporting line at $u$, then
\begin{equation}
s(u)<\inf_\alpha \{\alpha u-\varphi _g(\alpha )\}+g(u).
\end{equation}
\end{theorem}

This result effectively corrects for the nonequivalence of the
microcanonical and canonical ensembles, for it shows that, in cases where $s$
does not have a supporting line at $u$, we may be able to find a function
$g\neq 0$ that locally transforms $s(u)$ to a deformed entropy $s_g=s-g$
that has a supporting line at $u$. This induced supporting line property is
what enables use to write $s_g(u)$ as the LF transform of the deformed free
energy function $\varphi _g(\alpha )$, and, from there, we recover $s(u)$ by
simply adding $g(u)$ to the result of the LF transform of $\varphi _g(\alpha)$,
thereby undoing the deformation induced by $g$. In this case, we can
say, in parallel with was said in the previous section, that we have
\textit{equivalence of the microcanonical and generalized canonical ensembles at the
thermodynamic level}. Obviously, if $s_g$ does not possess a supporting line
at $u$ for the chosen $g$, then $s_g^{**}(u)\neq s_g(u)$, and so the trick
of expressing $s(u)$ through the LF transform of $\varphi _g(\alpha )$ does
not work. In this latter case, we say that there is \textit{thermodynamic
nonequivalence of the microcanonical and generalized canonical ensembles}.

\begin{figure*}[t]
\includegraphics{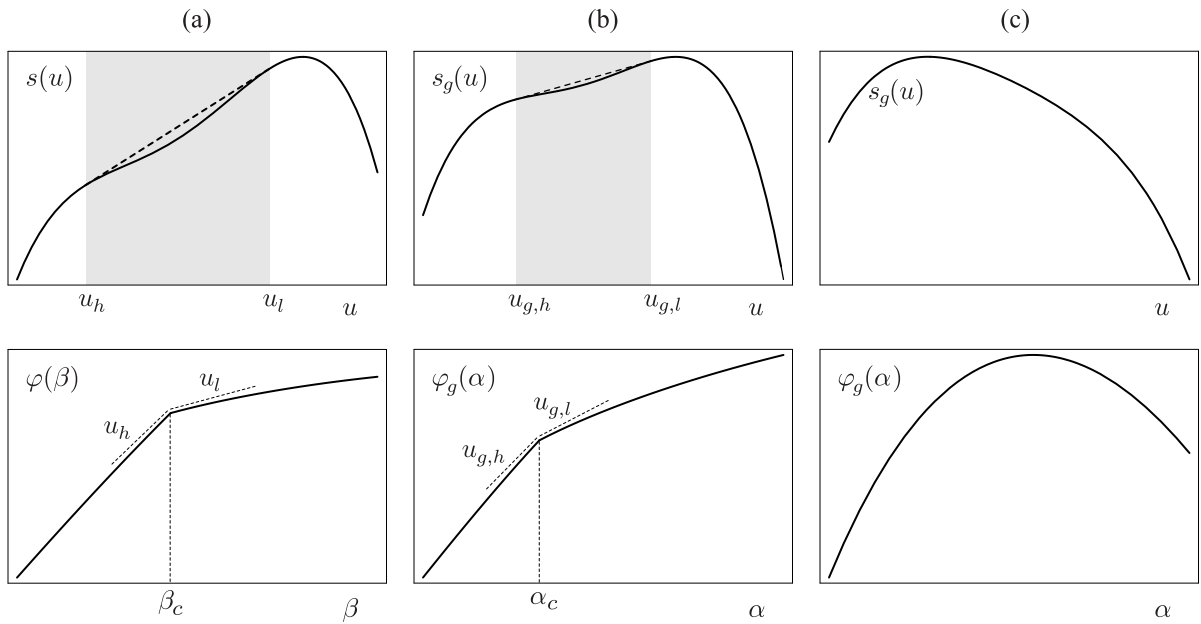}
\caption{Schematic illustration of the effect of $g$ on the entropy and free energy
functions.
(Left) Initial entropy $s(u)$ and its corresponding free energy
$\protect\varphi (\protect\beta)$ (see Figs.~\ref{figline} and \ref{figphi}).
(Middle) Modified entropy
$s_g(u)$ having a smaller region of nonconcavity than $s(u)$, and
its corresponding generalized free energy
$\protect\varphi_g (\protect\alpha)$. (Right) A modified entropy
$s_g(u)$ rendered fully concave by $g$;
its corresponding generalized free energy
$\protect\varphi_g (\protect\alpha)$ is everywhere differentiable.}
\label{figsphi}
\end{figure*}

We close our discussion of thermodynamic nonequivalence of ensembles by
stating the generalization of Theorem~\ref{Tphi}. We omit the proof of
this generalization as it follows directly
from well-known properties of LF transforms and a straightforward
generalization of well-known results about the equilibrium properties of the
canonical ensemble.

\begin{theorem}
Let $\varphi _g(\beta )$ denote the generalized free energy function defined
in $\emph{(\ref{genfe1})}$.

\emph{(a)} If $\varphi _g$ is differentiable at $\alpha $, then
\begin{eqnarray}
s(u_{g,\alpha }) &=&\varphi _g^{*}(u_{g,\alpha })+g(u_{g,\alpha })  \nonumber
\\
&=&\alpha u_{g,\alpha }-\varphi _g(\alpha )+g(u_{g,\alpha }),
\end{eqnarray}
where $u_{g,\alpha }=\varphi _g^{\prime }(\alpha )$ represents the
equilibrium value of $h$ in the generalized canonical ensemble with
parameters $\alpha $ and $g$.

\emph{(b)} if $\varphi _g$ is everywhere differentiable, then $s=\varphi
_g^{*}+g$ for all $u$ in the range of $h$.
\end{theorem}

The implications of this theorem are illustrated in Fig.~\ref{figsphi},
which shows the plots of different entropy and free energy
functions resulting from different choices for the function $g$. This figure
depicts three possible scenarios:

(a) The original nonconcave entropy function $s(u)$ and its associated
nondifferentiable free energy function $\varphi (\beta )$ for $g=0$. Recall
in this case that the extent of the nonconcave region of $s(u)$ is equal to
the latent heat associated with the nondifferentiable point of $\varphi
(\beta )$; see Fig.~\ref{figsphi}.

(b) The modified entropy function $s_g(u)$ resulting from this choice of $g$
has a smaller region of nonconcavity than $s(u)$, which is to say that
\begin{equation}
\Delta u_g=u_{g,h}-u_{g,l}<\Delta u.
\end{equation}
From the point of view of the generalized canonical ensemble, we have
\begin{equation}
\Delta u_g=\varphi _g^{\prime }(\alpha _c-0)-\varphi _g^{\prime }(\alpha
_c+0),
\end{equation}
and so we see that this choice of $g$ brings, in effect, the left- and
right- derivative of $\varphi _g$ at $\alpha _c$ closer to one another
compared to the case where $g=0$. In other words, this choice of $g$ has the
effect of ``inhibiting'' the first-order phase transition of the canonical
ensemble.

(c) There is a function $g$ that makes $s_g(u)$ strictly concave everywhere.
In this case, $\varphi _g(\beta )$ is everywhere differentiable, which means
that the first-order phase transition of the canonical ensemble has been
completely obliterated. Thus, by varying $\alpha $, it is now possible to
``scan'' with $u_{g,\alpha }$ any values of the mean Hamiltonian $h$, which
is a formal way to say that the generalized canonical ensemble can be used
to access any particular mean energy value of the microcanonical ensemble,
and so that both ensembles are equivalent.

\section{Macrostate nonequivalence of ensembles}

\label{Smacro}

Just as the thermodynamic properties of systems can generally be related to
their macrostates equilibrium properties, it is possible to define the
equivalence or nonequivalence of the microcanonical and canonical ensembles
at the macrostate level and relate this level to the thermodynamic level of
nonequivalent ensembles described earlier. This was done recently by Ellis,
Haven and Turkington \cite{ellis2000}. A full discussion of the results
derived by these authors would fill too much space; we will limit ourselves
here to present a summary version of the most important results found in
Ref.~\cite{ellis2000} and then present generalizations of these results which are
obtained by replacing the canonical ensemble with the generalized canonical
ensemble \cite{costeniuc2004}.

We first recall the basis for defining nonequivalent ensembles at the
macrostate level. Given a macrostate or order parameter $m$, we proceed to
calculate the equilibrium, that is, most probable values of $m$ in the
microcanonical and canonical ensembles as a function of the mean energy $u$
and inverse temperature $\beta $, respectively. Let us denote the first set
of microcanonical equilibrium values of $m$ parameterized as a function of
$u $ by $\mathcal{E}^u$ and the second set of canonical equilibrium values
parameterized as a function of $\beta $ by $\mathcal{E}_\beta $. By
comparing these sets, we then define the following. On the one hand, we say
that the microcanonical and canonical ensembles are \textit{equivalent at
the macrostate level} whenever, for a given $u$, there exists $\beta $ such
that $\mathcal{E}^u=\mathcal{E}_\beta $. On the other hand, we say that the
two ensembles are \textit{nonequivalent at the macrostate level} if for a
given $u$, there is no overlap between $\mathcal{E}^u$ and all possible sets
$\mathcal{E}_\beta $, that is, mathematically if $\mathcal{E}^u\cap
\mathcal{E}_\beta =\emptyset $ for all $\beta $.

These definitions of the macrostate level of equivalent and nonequivalent
ensembles can be found implicitly in the work of Eyink and Spohn
\cite{eyink1993}. They are stated explicitly in the comprehensive study of Ellis,
Haven and Turkington \cite{ellis2000}, who have proved that the
microcanonical and canonical ensembles are equivalent (resp., nonequivalent)
at the macrostate level when they are equivalent (resp., nonequivalent) at
the thermodynamic level. The main assumption underlying their work is that
the mean Hamiltonian function $h(\omega )$ can be expressed as a function of
the macrostate variable $m$ in the asymptotic limit where $n\rightarrow
\infty $. A summary of their main results is presented next; see
Ref.~\cite{ellis2000} for more complete and general results.

\begin{theorem}
\label{Tmac}
We say that $s$ admits a \emph{strict} supporting line at $u$ if
there exists $\beta $ such that $s(v)<s(u)+\beta (v-u)$ for all $v\neq u$.

\emph{(a)} If $s$ admits a strict supporting line at $u$, then $\mathcal{E}
^u=\mathcal{E}_\beta $ for some $\beta \in \Bbb{R}$, which equals
$s^{\prime}(u)$ if $s$ is differentiable at $u$.

\emph{(b)} If $s$ admits no supporting line at $u$, that is, equivalently,
if $s(u)\neq s^{**}(u)$, then $\mathcal{E}^u\cap \mathcal{E}_\beta
=\emptyset $ for all $\beta \in \Bbb{R}$.
\end{theorem}

The first case corresponds, as was stated above, to macrostate equivalence
of ensembles, whereas the second corresponds to macrostate nonequivalence of
ensembles. There is a third possible relationship that we omit from our
analysis because of too many technicalities involved: it is referred to as
\textit{partial equivalence} and arises when $s$ possesses a
\textit{non-strict} supporting line at $u$, that is, a supporting line that touches
the graph of $s(u)$ at more than one point \cite{ellis2000}.

Our next result is the generalization of Theorem \ref{Tmac} about macrostate
equivalence and nonequivalence of ensembles. It shows, in analogy with the
thermodynamic level, that the microcanonical properties of a system can be
calculated from the point of view of the generalized canonical ensemble when
the canonical ensemble cannot be used for that goal.

\begin{theorem}
\label{Tgenmac}
Let $s_g(u)=s(u)-g(u)$, where $g(u)$ is any continuous
function of the mean energy $u$, and let $\mathcal{E}_{g,\alpha }$ denote
the set of equilibrium values of the macrostate $m$ in the generalized
canonical ensemble with function $g$ and parameter $\alpha $.

$\emph{(a)}$ If $s_g$ admits a strict supporting line at $u$, then
$\mathcal{E}^u=\mathcal{E}_{g,\alpha }$ for some $\alpha \in \Bbb{R}$,
which equals $s_g^{\prime }(u)$ if $s_g$ is differentiable at $u$.

\emph{(b)} If $s_g$ does not admit a supporting line at $u$, that is,
equivalently, if $s_g(u)\neq s_g^{**}(u)$, then $\mathcal{E}^u\cap
\mathcal{E}_{g,\alpha }=\emptyset $ for all $\alpha \in \Bbb{R}$.
\end{theorem}

\proof
For the purpose of proving this result, we define a generalized
microcanonical ensemble by changing the Lebesgue measure $\mu (\omega)
=d\omega $, which underlies the definition of the microcanonical ensemble,
to the measure
\begin{equation}
\mu _g(\omega )=e^{-ng(h(\omega ))}d\omega .
\end{equation}
As mentioned before, the extra factor $e^{-ng(h(\omega ))}$ modifies the
microcanonical entropy $s(u)$ to $s_g(u)$ as shown in (\ref{defent1});
however, and this is a crucial observation, it leaves all the macrostate
equilibrium properties of the microcanonical ensemble unchanged because
\textit{all the microstates that have the same mean energy still have the
same weight}. This implies that the generalized microcanonical ensemble is,
by construction, always equivalent to the microcanonical ensemble at the
macrostate level. That is to say, if $\mathcal{E}_g^u$ denotes the set of
equilibrium values of the macrostate $m$ with respect to the generalized
microcanonical ensemble with mean energy $u$ and function $g$, then
$\mathcal{E}_g^u=\mathcal{E}^u$ for all $u$ and all $g$.

Next we observe that the supporting line properties of $s_g(u)$ determine
whether the generalized microcanonical and generalized canonical ensembles
are equivalent or not, just as the supporting line properties of $s(u)$
determine whether or not the standard microcanonical and standard canonical
ensembles are equivalent; to be sure, compare equations (\ref{lf1}) and
(\ref{genlf1}).

With these two observations in hand, we are now ready to prove equivalence
and nonequivalence results between $\mathcal{E}^u$ and $\mathcal{E}
_{g,\alpha }$. Indeed, all we have to do is to use the equivalence and
nonequivalence results of Theorem~\ref{Tmac} to first derive equivalence
and nonequivalence results about
$\mathcal{E}_g^u$ and $\mathcal{E}_{g,\alpha }$, and then transform these to
equivalence and nonequivalence results between $\mathcal{E}^u$and
$\mathcal{E}_{g,\alpha }$ using the fact that $\mathcal{E}^u=\mathcal{E}_g^u$ for all
$u $ and any choice of $g$. To prove Part (a), for example, we reason as
follows. If $s_g$ admits a strict supporting line at $u$, then $\mathcal{E}
_g^u=\mathcal{E}_{g,\alpha }$ for some $\alpha \in \Bbb{R}$. But since
$\mathcal{E}_g^u=\mathcal{E}^u$ for all $u$ and any $g$, we obtain
$\mathcal{E}^u=\mathcal{E}_{g,\alpha }$ for the same value of $\alpha $.
Part (b) is proved similarly. If $s_g$ admits no supporting line at $u$, that is, if
$s_g(u)\neq s_g^{**}(u)$, then $\mathcal{E}_g^u\cap \mathcal{E}_
{g,\alpha}=\emptyset $ for all $\alpha \in \Bbb{R}$. Using again the equality
$\mathcal{E}_g^u=\mathcal{E}^u$, we thus obtain $\mathcal{E}^u\cap \mathcal{E}_
{g,\alpha }=\emptyset $ for all $\alpha \in \Bbb{R}$.
\endproof

\section{Gaussian ensemble}

\label{Sgauss}

The choice $g(u)=\gamma u^2$ defines an interesting form of the generalized
canonical ensemble that was introduced more than a decade ago by
Hetherington \cite{heth1987} under the name of \textit{Gaussian ensemble};
see also Refs.~\cite{stump1987,challa1988,challa21988,kiessling1997,johal2003}.
Many properties of this ensemble were studied by Challa and Hetherington
\cite{challa1988,challa21988} who showed, among other things, that the
Gaussian ensemble can be thought of as arising when a sample system is put
in contact with a finite heat reservoir. From this point of view, the
Gaussian ensemble can be thought of as a kind of ``bridge ensemble'' that
interpolates between the microcanonical ensemble, whose definition involves
no reservoir, and the canonical ensemble, whose definition involves an
infinite reservoir.

The results presented in this paper imply a somewhat different
interpretation of the Gaussian ensemble. They show that the Gaussian
ensemble can in fact be made equivalent with the microcanonical ensemble, in
the thermodynamic limit, when the canonical ensemble cannot. A trivial
implication of this is that the Gaussian ensemble can also be made
equivalent with both the microcanonical and canonical ensembles if these are
already equivalent. The precise formulation of these equivalence results is
contained in Theorems~\ref{Tgenth} and \ref{Tgenmac} in which $s_g(u)$
takes the form $s_\gamma(u)=s(u)-\gamma u^2$.

\begin{figure}[t]
\includegraphics{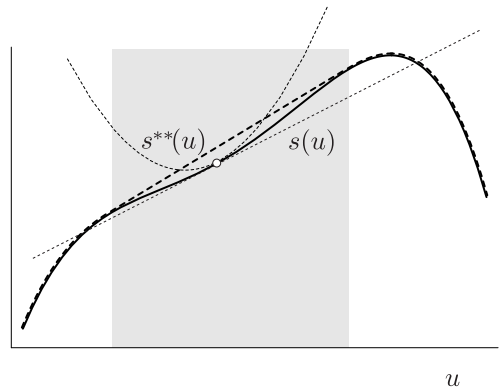}
\caption{Example of a point of $s(u)$ which does not admit a supporting line
but admits a supporting
parabola. Such a point is accessible to the Gaussian ensemble but not
to the canonical ensemble.}
\label{figpara}
\end{figure}

In the specific case of the Gaussian ensemble, these results can be
rephrased in a more geometric fashion using the fact that a supporting line
condition for $s_\gamma $ at $u$ is equivalent to a supporting parabola
condition for $s$ at $u$. To see this, we need to substitute the expression
of $s_\gamma (u)$ and $\alpha =s_\gamma ^{\prime }(u)=s^{\prime }(u)-2\gamma
u$ in the definition of the supporting line to obtain
\begin{equation}
s(v)\leq s(u)+\alpha (v-u)+\gamma (v-u)^2
\end{equation}
for all $v$. We assume at this point that $s_\gamma $, and therefore $s$,
are differentiable functions at $u$. The right-hand side of this inequality
represents the equation of a parabola that touches the graph of $s$ at $u$
and lies above that graph at all other points (Fig.~\ref{figpara});
hence the term supporting parabola. As a result of this
observation, we then have the following: if $s$ admits a supporting parabola
at $u$ (Fig.~\ref{figpara}), then
\begin{eqnarray}
s(u) &=&\varphi _\gamma ^{*}(u)+\gamma u^2  \nonumber \\
&=&\inf_\alpha \{\alpha u-\varphi _\gamma (\alpha )\}+\gamma u^2;
\label{gausss1}
\end{eqnarray}
otherwise the above equation is not valid. A macrostate extension of this
result can be formulated in the same way by transforming the supporting line
condition for $s_\gamma $ in Theorem~\ref{Tgenmac} by a supporting parabola
condition for $s(u)$.

The advantage of using supporting parabola instead of supporting lines is
that many properties of the Gaussian ensemble can be proved in a simple,
geometric way. For example, it is clear that since $s(u)$ can possess a
supporting parabola while not possessing a supporting line (Fig.~\ref{figpara}),
the Gaussian ensemble does indeed go beyond the standard
canonical ensemble. Moreover, the range of nonconcavity of $s_g(u)$ should
shrink as one chooses larger and larger values of $\gamma $. From this last
observation, it should be expected that a single (finite) value of $\gamma $ can in
fact be used to achieve equivalence between the Gaussian and microcanonical
ensembles for all value $u$ in the range of $h$, provided that (i) $\gamma $
assumes a large enough value, basically greater that the largest second
derivative of $s(u)$; (ii) that the graph of $s(u)$ contains no corners,
that is, points where the derivative of $s(u)$ jumps and where $s^{\prime
\prime }(u)$ is undefined; see Ref.~\cite{costeniuc2004} for details.

The second point implies physically that the Gaussian ensemble
with $\gamma < \infty$ cannot be applied at points of first-order
phase transitions in the microcanonical ensemble. Such
points, however, can be dealt with within the Gaussian ensemble
by letting $\gamma\rightarrow\infty$, as we shall show in a forthcoming
paper \footnote{R.S. Ellis and H. Touchette, in preparation.}.
With the proviso that the limit $\gamma\rightarrow\infty$ may have to
be taken, we can then conclude that the Gaussian ensemble is
a universal ensemble: in theory, it can recover any shape of
microcanonical entropy function through Eq.(\ref{gausss1}), which
means that it can achieve equivalence with the microcanonical
ensemble for any system.

\section{Conclusion}

\label{Scon}

In this paper we have studied a generalization of the canonical ensemble
which can be used to assess the microcanonical equilibrium properties of a
system when the canonical ensemble is unavailing in that respect because of
the presence of nonconcave anomalies in the microcanonical entropy function.
Starting with the supporting properties of the microcanonical entropy, which
are known to determine the equivalence and nonequivalence of the
microcanonical and canonical ensembles, we have demonstrated how these
properties can be extended at the level of a modified form of the
microcanonical entropy to determine whether the microcanonical and
generalized canonical ensembles are equivalent or not.
Equivalence-of-ensembles conditions for these two ensembles were also given
in terms of a generalized form of the canonical free energy. Finally, we
have discussed the case of the Gaussian ensemble, a statistical-mechanical
ensemble introduced some time ago by Hetherington, which arises here as a
specific instance of our generalized canonical ensemble. 
For the Gaussian ensemble, results establishing the
equivalence and nonequivalence with the microcanonical
ensemble were given in terms of supporting parabolas.

In forthcoming papers, we will present applications of the generalized
canonical ensemble for two simple spin models which are known to possess a
nonconcave microcanonical entropy function. The first one is the
Curie-Weiss-Potts model studied in Refs.~\cite{ispolatov2000,costeniuc2005}; the
second is the block spin model studied in Refs.~\cite{touchette22003,touchette22005}.

\begin{acknowledgments}

The research of M.C. and R.S.E. was supported by a grant from the National
Science Foundation (NSF-DMS-0202309); that of B.T. was supported by a grant
from the National Science Foundation (NSF-DMS-0207064). H.T. was supported
by the Natural Sciences and Engineering Research Council of Canada and the
Royal Society of London (Canada-UK Millennium Fellowship).

\end{acknowledgments}

\bibliography{gencanv2}

\end{document}